\documentclass[12pt]{iopart}
\usepackage{graphicx}
\begin{document}

%\begin{titlepage}
\title[$O(N)$ TBMD Simulation of Thermal Stability of SWCNT ]{Thermal Stability of Metallic Single-Walled Carbon Nanotubes:
An O(N) Tight-Binding Molecular Dynamics Simulation Study}
\author{G. Dereli$^1$\footnote{Corresponding author: gdereli@yildiz.edu.tr} ,
B. S\"{u}ng\"{u}$^1$, C. \"{O}zdo\u{g}an$^2$ }

\address{$^1$ Department of Physics, Y{\i}ld{\i}z Technical
University, 34210 Istanbul, Turkey}
\address{$^2$ Department of Computer Engineering, \c{C}ankaya University, 06530 Ankara, Turkey}

\begin{abstract}

\noindent Order(N) Tight-Binding Molecular Dynamics (TBMD)
simulations are performed to investigate the thermal stability of
$(10,10)$ metallic Single-Walled Carbon Nanotubes (SWCNT). Periodic
boundary conditions (PBC) are applied in axial direction. Velocity
Verlet algorithm along with the canonical ensemble molecular
dynamics (NVT) is used to simulate the tubes at the targeted
temperatures. The effects of slow and rapid temperature increases on
the physical characteristics, structural stability and the
energetics of the tube are investigated and compared. Simulations
are carried out starting from room temperature and the temperature
is raised in steps of 300K. Stability of the simulated metallic
SWCNT is examined at each step before it is heated to higher
temperatures. First indication of structural deformation is observed
at 600K. For higher heat treatments the deformations are more
pronounced and the bond breaking temperature is reached around
2500K. Gradual (slow) heating and thermal equilibrium (fast heating)
methods give the value of radial thermal expansion coefficient  in
the temperature range between 300K-600K as $0.31 \times
10^{-5}$(1/K) and $0.089 \times 10^{-5}$(1/K), respectively. After
600K, both methods give the same value of $0.089 \times
10^{-5}$(1/K). The ratio of the total energy per atom with respect
to temperature is found to be $3 \times 10^{-4}$eV/K.
\end{abstract}


\maketitle
%\end{titlepage}

\section{Introduction}

\noindent  There has been increasing attention on nanotechnological
research due to the demand for the most durable nanoscale structures
in recent years. SWCNTs are in the first group of materials in this
respect due to their high tensile strength and adjustable electrical
conductivity.  SWCNTs high thermal and chemical stability can
conveniently be applied to gas sensors, dielectric devices,
nanoelectronic devices, nanocomposites and emitters. SWCNTs high
thermal stability can be compared with other single walled
nanotubes. For example, Single Walled Aluminum Nitride Nanotubes
(SWAlNNTs) can stabily exist at room temperature but start to melt
when temperature is higher than 600 K [1]. This tendency to deform
easily in the radial direction is a disadvantage for possible
applications. On the other hand, studies on the high temperature
stability of gold nanotubes show that they are barely stable up to
1200K with total energy per atom $2.8 \times 10^{-4}$eV/K [2,3].
Therefore there is an increasing attention on the thermal behavior
of metallic CNTs which varies  depending on their diameters, lengths
and chiralities.

\medskip

\noindent The available experimental and theoretical studies on
thermal stability of SWCNTs show that they are still in an initial
stage. The available results are limited and not in agreement with
each other. A $4{\AA}$ diameter nanotube having a single vacancy
with three dangling bonds has been found to retain its cylindrical
shape at high temperatures around 4000 K [4]. Liew et al. [5]
investigated the thermal properties of various CNTs using MD
simulation method. They found that SWCNTs are thermally more stable
than MWCNTs. They also showed that shorter CNTs are able to
withstand higher thermal loads. CNTs with a larger diameter are also
more resistant to thermal loads. On the other hand temperature
affects the structural properties and energetics of CNTs. The
coalescence of SWCNTs is investigated by Terrones et al. [6] using
not only in situ under electron irradiation at high temperatures in
TEM, but also by TBMD and Monte Carlo simulations. They observed
that SWCNTs heated above $2000 C^{0}$ in Ar or He atmospheres lead
to either MWCNTs or coalesced SWCNTs of larger diameters. Metenier
et al. [7] investigated the effect of thermal treatment under argon
gas flow on the evolution of SWCNT bundles. Their experiments showed
that the SWCNT bundles are not affected up to $1600 C^0$, however,
 coalescence of SWCNTs starts to occur between $1800-2000 C^0$.
From the heat treatment at $2200 C^0$, the disappearance of SWCNTs
to the benefit of MWCNTs is observed. Same results are also observed
in the experimental study by Yudasaka et al. [8].  In accord with
these results  an MD simulation is performed by Lopez et al. [9].
They showed that bundles of SWCNTs are stable under thermal
treatment up to $1600 C^0$. Above $2200 C^0$  they become unstable
and transform into MWCNTs. The effect of heat treatment on thermal
stability and structural changes of DWCNTs is investigated by Kim et
al. [10]. They showed that DWCNTs are structurally stable up to
$2000 C^0$. Between $2100-2400 C^0$ the outer walls of adjacent
DWNTs start coalescing into large diameter tubes. Between $2500-2800
C^0$ MWNTs and flaky carbons were observed using experimental
methods. Kawai et al. [11] also investigated the coalescence of
Ultrathin Carbon Nanotubes (UTCNTs) using TBMD simulation technique.
They have found that above $2000 C^0$, two UTCNTs having either the
same chirality or different chiralities can coalesce without
initially introducing atomic defects to enhance the reaction.

\medskip

\noindent  In this paper, we study the thermal history of a (10,10)
SWCNT using our O(N), TBMD approach which previously proved to be
successful in simulations of electronic structure and elastic
properties of the same nanotube [12,13]. The effects of slow and
rapid temperature increases on the physical characteristics,
structural stability and the energetics of the tube are investigated
and compared.

\section{Method}

\noindent The traditional TB solves the Schr\"{o}dinger equation in
reciprocal space by direct matrix diagonalization which results in
cubic scaling with respect to the number of atoms. The O(N) methods
on the other hand  solve for the band energy in real space and make
the approximation that only the local environment contributes to the
bonding, and hence band energy, of each atom. In fact all the O(N)
methods in which the properties of the whole system are computed
such as the total energy  or the forces on all atoms necessarily
involve approximations to the exact solution of the effective
one-electron Hamiltonian. These approximations are based on physical
assumptions that are generally connected to the locality or
near-sightedness principle in one way or another. The O(N) scaling
arises from the decay and/or truncation of certain quantities. We
have used TBMD algorithms involving an energy functional and a
parametrization that was previously proven successful in heat
applications [6]. In a previous work two of the authors (G. D. and
C. \"{O}.) have improved and successfully applied the Order(N)
techniques to these TBMD algorithms in simulations of SWCNTs.
Details of the technique and the parametrization can be followed in
[12-14] and the references therein.

\medskip

\noindent In the present  work a (10,10) single walled CNT
consisting of 400 atoms with 20 layers is simulated using $O(N)$
parallel tight binding molecular dynamics algorithms.  In the
implementation of the $O(N)$ technique we adopted a Divide and
Conquer approach. The accuracy of the description is enhanced by the
use of basis functions of only the neighboring atoms that is called
the "buffer". The Schr\"{o}dinger equation of the buffer has the
same form as in [14]. The eigenvalues and eigenvectors are found by
diagonalizing the Hamilton matrix for  each subsystem.  Buffer size
and the cuboidal box size (also called DAC box size) are the two
important parameters of the $O(N)$ algorithms. We took the DAC box
size equal to the distance between two consecutive cross-sectional
layers  ($1.229{\AA}$) along the uniaxial direction in a $(10,10)$
tube [12].  This provides the same number of interacting neighbor
atoms for each subsystem. Periodic boundary condition is applied in
the uniaxial direction. All the simulations presented here are
carried out in the canonical (N,V,T) ensemble. The Newtonian
equations of motion are integrated using the Velocity Verlet
algorithm with a time step equal to 1 fs. To avoid an inaccurate
integration, the velocities of the constituent atoms are
occasionally rescaled to maintain the temperature of the system at
the target value. The difference of O($N^3$) total energy result
with $O(N)$ total energy result is indicated as error. By tuning the
above two parameters we try to minimize the error. The errors for
different buffer sizes are found for the $(10,10)$ tube and the
smallest error is read at the buffer size of $4.8{\AA}$ [15].  The
time step of the simulations determines the real time of the
simulation. Before starting the production phase of the simulations
careful study of the time step is done. A time step that brings the
system quickly to thermal equilibrium is chosen as 1 fs. It is also
made sure that the possibility of the system getting trapped in a
metastable state is avoided. Cut-offs for the interactions and the
bond length, bond angle distribution functions are $2.6{\AA}$ and
$2.1{\AA}$, respectively. It is known that temperature can be
related to the average energy of a system of particles in
equilibrium. This definition also works for nanoscale systems such
as CNTs. MD simulations calculate the position  and the velocity  of
each atom at each time step. We store the velocity values and
compute the average kinetic energy over N steps in time:
\begin{equation}
<K.E.> = \frac{m}{2N} \sum_{n=1}^{N} {v_i}^2(t_n) =
<\frac{1}{2}m{v_i}^2> \quad .
\end{equation}
Here the averaging must be done over very long times in order to
obtain good statistical average kinetic energies.  Then kinetic
energy can be converted into a temperature scale using
\begin{equation}
 <\frac{1}{2}m{v_i}^2> = \frac{3}{2} k_B T_i \quad .
\end{equation}
Thus the simulation temperature is controlled by rescaling the
velocities.

\section{Results and Discussion}

\noindent During our simulations the SWCNTs were heated by two
different methods. In the first method which is based on "gradual
(slow) heating", the system has been brought to equilibrium at 300K
temperature during a 5 ps of run. Thermal history applied to the
SWCNT can be followed in Figure 1.
\begin{figure}
\centering
\includegraphics{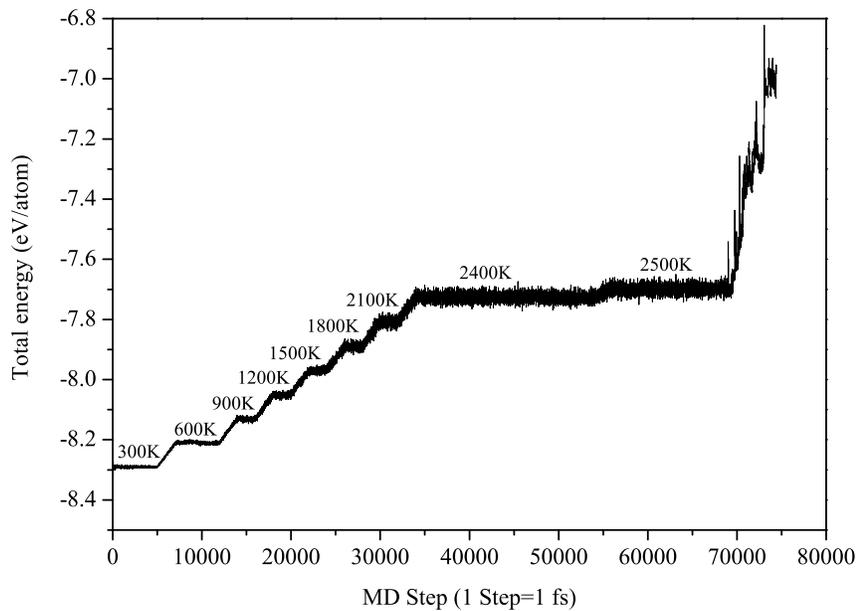}
\caption{Thermal history of total energy per atom of (10,10) SWCNT
as a function of simulation time in gradual heating. Tube decays at
2500K.}
\end{figure}
Temperature has been increased in steps of 300K. System has been
brought to equilibrium in 2 ps periods at each targeted temperature.
Information about the strain energy of the tube at each temperature
can be followed from this figure. As the temperature increased
longer periods of relaxations are applied in order to observe the
detachment of atoms from the $(10,10)$ SWCNT.  Bond breaking
temperature is observed at 2500K. At this temperature, after 13 ps
of relaxation one atom is detached and as the simulation proceeds,
the number of detached atoms increased and a tearing effect is
observed [15]. In Figure 1, sharp peaks after 13 ps of relaxations
indicate the detachment of atoms. Starting from 600K, hexagons are
deformed but carbon atoms kept their bonding until 2500K. In Figure
2 we present the change of total energy per atom with temperature.
Kinetic energy, band-structure energy and the repulsive potential
energy contributions to the total energy are also displayed in
Figure 2.
\begin{figure}
\centering
\includegraphics{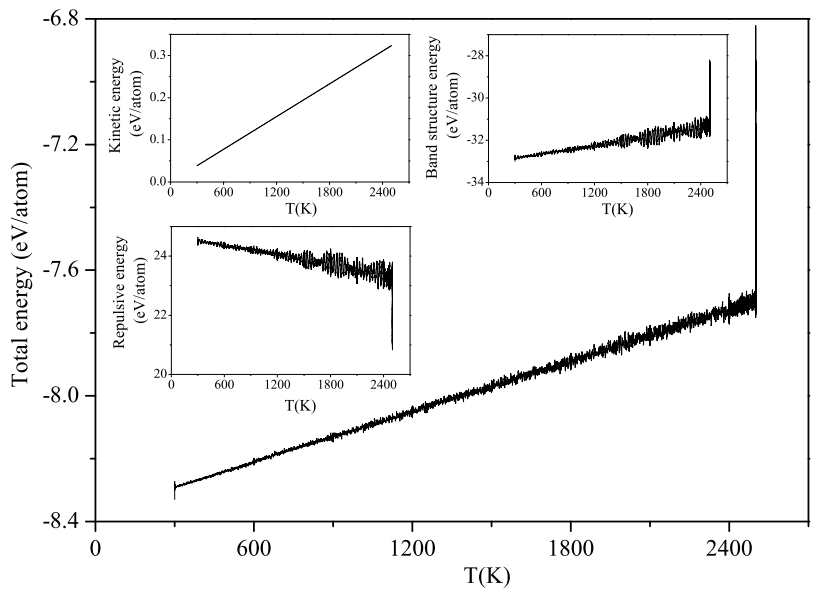}
\caption{Total energy per atom as a function of temperature. Kinetic
energy, band-structure energy and the repulsive potential energy
contributions of the total energy are given. }
\end{figure}
Kinetic energy and band-structure energy increase with temperature
as expected. Repulsive potential energy decreases with temperature.
The variation of kinetic energy as a function of temperature yields
a fixed value $1.3 \times 10^{-4}$eV/K that is in accord with the
equipartition theorem (Eq.2). As a result total energy increases as
the temperature rises and the variation of energy as a function of
temperature yields $3 \times 10^{-4}$eV/K. This is in agreement with
the only available result in literature for fullerenes [16].

\noindent In Figure 3, we show the changes in the physical
properties of our SWCNTs change temperature that are displayed
through the radial distribution functions, bond-length and
bond-angle distribution functions, respectively. .
\begin{figure}
\centering
\includegraphics{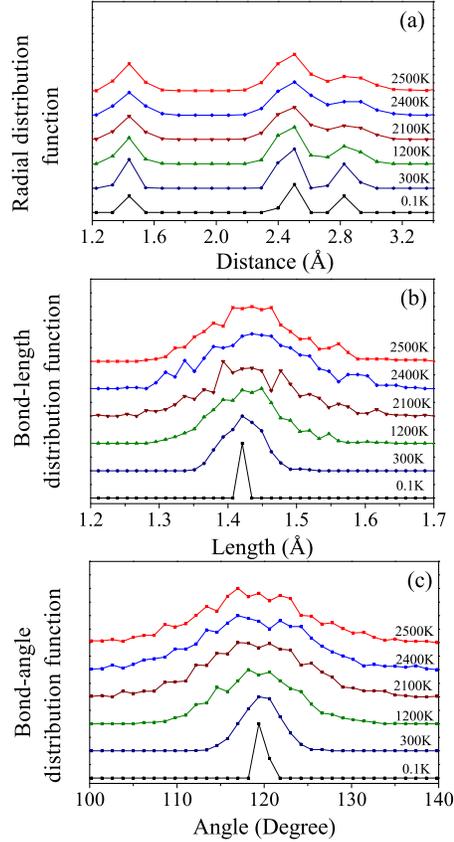}
\caption{Physical properties obtained in gradual heating: (a) Radial
distribution functions, (b) Bond-length distribution functions, c)
Bond-angle distribution functions.}
\end{figure}
In Figure 3(a) at 300K, we observe the first peak at the nearest
neighbor distance of  $1.44{\AA}$, the second peak is at the second
nearest neighbor distance of $2.50{\AA}$ and the third peak at
$2.82{\AA}$. These values are all in accord with the observed
properties of $(10,10)$ tubes [17]. As the temperature is increased
a broadening of the first peak is observed. Second peak is deformed
and the third peak is almost lost, suggesting that the crystalline
phase has been transformed to the amorphous phase. Bond-length
distribution function peaks around $1.42{\AA}$ which is the
$a_{c-c}$ distance for the $(10,10)$ tube at 300K. The temperature
increase affects the peaks that are broadened and do not remain
sharp above 1200K. Hexagons of carbon atoms are deformed largely
until the bond-breaking temperature of 2500K. In Figure 3(b) the
peak at 2500K is the bond length distribution obtained before the
detachment of atoms begin. Bond-angle distribution function peaks at
119.4 degree at 300K. As the temperature increases peak positions
are shifted and peaks are broadened. Above 1200K peaks are no longer
sharp.
\begin{figure}
\centering
\includegraphics{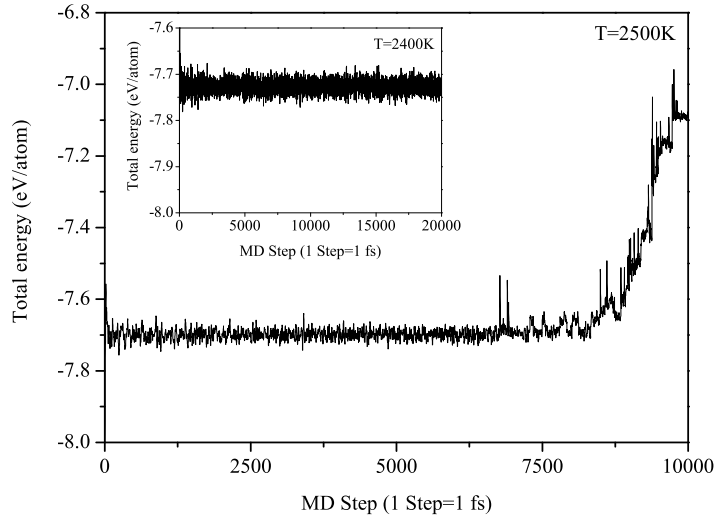}
\caption{ (a) Total energy per atom of the optimized (10,10) SWCNT
at 2400K and (b) at 2500K. }
\end{figure}
\noindent During our simulations we also used a fast heating method
that we call "thermal equilibrium method" to check the bond-breaking
mechanism of (10,10) SWCNT with temperature increase. In this
method, (10,10) SWCNT ought to  be optimized at discrete temperature
values. We chose the same target temperatures as in gradual (slow)
heating. Optimized tubes are left to reach thermal equilibrium at
these temperature values. Figure 4 shows that  the tube is stable at
2400K.  However, the optimized tube left in thermal equilibrium at
2500K  is not stable  after 7 ps of equilibration time. At the same
temperature with gradual heating, after 7 ps of equilibration time
atoms start to get detached and as the simulation proceeds the
number of detached atoms increase.

\noindent We calculated the average radius of the (10,10) SWCNT as
$6.785{\AA}$ at 300K in accordance with literature. Average radius
increases with temperature. (Nanotube length will also change
accordingly since we are using NVT algorithms.)  In Figure 5 we
present the average diameter enlargement with temperature using both
heating methods.
\begin{figure}
\centering
\includegraphics{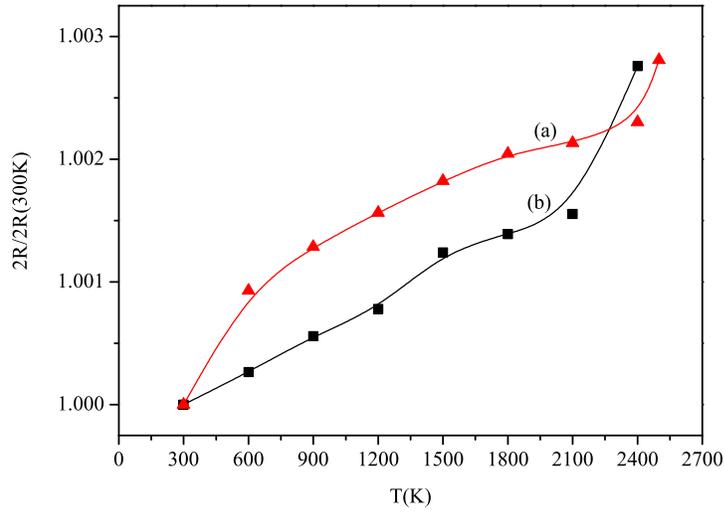}
\caption{Diameter enlargement with temperature: (a) gradual (slow)
heating method, b) thermal equilibrium (fast heating) method. }
\end{figure}
Radial thermal expansion coefficient of (10,10) SWCNT is calculated
from the linear regions in the temperature range between 300-600 K
of these patterns. Using Figure 5(a) radial thermal expansion
coefficient is calculated as $0.31 \times 10^{-5}$(1/K). This is the
result of gradual (slow) heating method. Using Figure 5(b) we
calculated this value as $0.089 \times 10^{-5}$(1/K). This is the
result of thermal equilibrium (fast heating) method. The
temperature-dependency of the  thermal expansion indicated in Figure
5 may be shown by taking the derivative of radius-temperature
curves. The temperature-dependency of average radius change is
illustrated in Figure 6.
\begin{figure}
\centering
\includegraphics{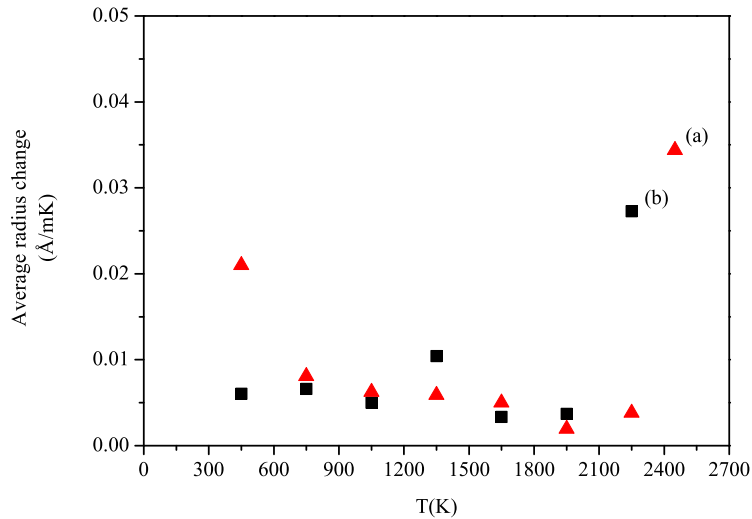}
\caption{The temperature dependency of average radius change: (a)
gradual (slow) heating method, (b) thermal equilibrium (fast
heating) method. }
\end{figure}
\noindent In the literature, several groups have reported
contradicting experimental and MD simulation results for the radial
thermal expansion of CNTs [18-21]. Maniwa et al. [18] performed
X-ray diffraction (XRD) studies for the thermal expansion of SWCNT
bundles. They determined the radial thermal expansion coefficient as
$(-0.15 \pm 0.20) \times 10^{-5}$(1/K) for SWCNT bundles. On the
other hand, the experimental and MD simulation results of Raravikar
et al. [19] calculated  the radial thermal expansion coefficient as
$0.08 \times 10^{-5}$(1/K). Using a different MD simulation method
Schelling et al. [20] also reports the same value. Negative radial
thermal expansion coefficient of [21] is commented on by [22-24].
Our radial thermal expansion coefficient value found from the
"gradual (slow) heating" method gives a value consistent  with [18].
On the other hand "thermal equilibrium (fast heating)" value of
$0.089 \times 10^{-5}$(1/K) is exactly the same as given by [19,20].
This shows the importance of the choice of a heating procedure
between 300K and 600K. The temperature-dependency of average radius
change as illustrated in Figure 6 shows that after 600K, both
methods give the same value of $0.089 \times 10^{-5}$(1/K).

\section{Conclusion}

\noindent We investigated the thermal characteristics of (10,10)
SWCNT using our O(N) TBMD simulation method. First we focused on the
effects of  temperature change  on physical properties such as bond
angle, bond length and radial distribution functions; structural
stability and the energetics of the tube. We showed that the tube
was deformed with increasing temperature but sustained its
structural stability up to high temperatures around 2500K. Both the
kinetic energy and band structure energy increase while repulsive
potential energy decreases as the temperature rises. The variation
of kinetic energy as a function of temperature yields a fixed value
given by $1.3 \times 10^{-4}$eV/K in accordance with the
equipartition theorem (Eq.2). As a result total energy increases as
the temperature rises and the variation of energy as a function of
temperature yields $3 \times 10^{-4}$eV/K in agreement with Kim et
al.'s [16] results for fullerenes. At low temperatures we observed
that the bond angle and bond length between carbon atoms have the
values 119.40 degree and $1.42{\AA}$, respectively. These are in
accordance with the hexagonal lattice of SWCNTs made up of graphene
sheets. We also investigated the radial distribution functions of
the tube and determined that the first third neighboring atoms
locate at distances for a reference atom at $1.44{\AA}$,
$2.50{\AA}$, $2.82{\AA}$, respectively. With increasing temperature,
because of the thermal motions of atoms in the lattice we observed
that the atomic vibrations  lead to changes in bond-angle,
bond-lengths and radial distribution functions. Also from the
temperature dependence of radial distribution functions results, we
propose that the variation of the third nearest neighbor position
shows the thermal expansion of the tube in radial direction.  We
further determined the enlargement of the tube diameter with
increasing temperature. We observed an increase from 1.357 nm to
1.361 nm within the temperature range from 300K to 2400K, in
agreement with the experimental results of Yudasaka et al. [25].
Gradual (slow) heating and the thermal equilibrium (fast heating)
methods calculate the radial thermal expansion coefficient in the
temperature range between 300K-600K as $0.31 \times 10^{-5}$(1/K)
and $0.089 \times 10^{-5}$(1/K), respectively. As the temperature
increases, gradual (slow) heating result approaches the thermal
equilibrium (fast heating) result of $0.089 \times 10^{-5}$(1/K).
Our gradual (slow) heating simulations corresponds to thermal
processes performed using conventional hot-wall furnaces in which
temperatures are applied slowly. On the other hand thermal
equilibrium (fast heating) simulations bring SWCNTs to a high
temperature rapidly. The general trend in thermal processing is to
reduce the process temperature and duration as much as possible in
order to restrict the motion of atoms through atomic diffusion. Fast
thermal processing restricts the diffusion which is important when
the control of impurities in the process is important. Fast thermal
processes are important for semiconductor device technology.
Experimentally, non steady-state investigation methods has been
intensively applied to the study of melting of graphite. Fast
heating enables one to obtain equilibrium thermal properties such as
enthalpy of melting, melting heat as well as temperature. Additional
structural features that affect the measurement of melting
temperature especially in the vicinity of melting point requires
pulsed heating experiments. Pulsed laser heating of graphite shows
an absence of a melting temperature plateau in the heating of low
density graphite specimen. Carbon vapor, as a result of graphite
sublimation, usually plays a leading role in the temperature
measurements near the melting point under gradual heating [26].
Through our fast and slow heating simulations of SWCNTs  we draw
attention to this issue.

\section{Acknowledgement}

\noindent  The research reported here is supported through the
Y{\i}ld{\i}z Technical University Research Fund Project No:
24-01-01-04. The calculations are performed at the Carbon Nanotubes
Simulation Laboratory at the Department of Physics, Y{\i}ld{\i}z
Technical University, \.{I}stanbul, Turkey.

\end{document}